\documentclass[sigconf]{acmart}

\AtBeginDocument{%
  }

\copyrightyear{2025} 
\acmYear{2025} 
\setcopyright{acmlicensed}\acmConference[CHI '25]{CHI Conference on Human Factors in Computing Systems}{April 26-May 1, 2025}{Yokohama, Japan}
\acmBooktitle{CHI Conference on Human Factors in Computing Systems (CHI '25), April 26-May 1, 2025, Yokohama, Japan}
\acmDOI{10.1145/3706598.3713287}
\acmISBN{979-8-4007-1394-1/25/04}

\usepackage{xcolor}
\usepackage{soul}
\setulcolor{blue}

\newcommand*{\corecode}[1]{#1}
\newcommand*{\subcode}[1]{\textbf{\ul{#1}}}

\newcommand*{\supersubsection}[1]{\textsc{#1}}

\begin{document}

\title{Creative Writers’ Attitudes on Writing as Training Data for Large Language Models}

\author{Katy Ilonka Gero}
\email{katy@g.harvard.edu}
\affiliation{%
  \institution{Harvard University}
  \city{Cambridge}
  \state{Massachusetts}
  \country{USA}
}

\author{Meera Desai}
\email{madesai@umich.edu}
\affiliation{%
  \institution{University of Michigan}
  \city{Ann Arbor}
  \state{Michigan}
  \country{USA}
}

\author{Carly Schnitzler}
\email{cschnit1@jh.edu}
\affiliation{%
  \institution{Johns Hopkins University}
  \city{Baltimore}
  \state{Maryland}
  \country{USA}
  }

\author{Nayun Eom}
\email{neom@fas.harvard.edu}
\affiliation{%
  \institution{Harvard University}
  \city{Cambridge}
  \state{Massachusetts}
  \country{USA}
}

\author{Jack Cushman}
\email{jcushman@law.harvard.edu}
\affiliation{%
  \institution{Harvard University}
  \city{Cambridge}
  \state{Massachusetts}
  \country{USA}
}

\author{Elena L. Glassman}
\email{glassman@seas.harvard.edu}
\affiliation{%
  \institution{Harvard University}
  % \streetaddress{150 Western Ave}
  \city{Cambridge}
  \state{Massachusetts}
  \country{USA}
  % \postcode{02134}
}

\renewcommand{\shortauthors}{Gero et al.}

\begin{abstract}
The use of creative writing as training data for large language models (LLMs) is highly contentious and many writers have expressed outrage at the use of their work without consent or compensation. In this paper, we seek to understand how creative writers reason about the real or hypothetical use of their writing as training data. We interviewed 33 writers with variation across genre, method of publishing, degree of professionalization, and attitudes toward and engagement with LLMs. We report on core principles that writers express (support of the creative chain, respect for writers and writing, and the human element of creativity) and how these principles can be at odds with their realistic expectations of the world (a lack of control, industry-scale impacts, and interpretation of scale). Collectively these findings demonstrate that writers have a nuanced understanding of LLMs and are more concerned with power imbalances than the technology itself. 
\end{abstract}





\maketitle

\section{Introduction}

As large language models have gained wide-spread usership, so have investigations into the text used to train them \cite{chang2023speak,gururangan2022whose,luccioni2021s,dodge2021documenting,gehman2020realtoxicityprompts,bandy2021addressing,longpre2024large}. There has been a particularly swift and negative response to the use of creative writing, such as novels, memoirs, and longform journalistic books, as training data, as the authors of this writing express outrage at the use of their work without consent or compensation \cite{asmelash2023thesebooksused}. Such frustration has led to dozens of lawsuits against technology companies like OpenAI, Microsoft, and Google, alleging that this use of writing goes against U.S. copyright law \cite{genaicasetracker}. 

Not everyone agrees that the use of creative writing for training data is illegal, or even immoral. In comments to the U.S. copyright office, The Author’s Alliance, whose mission is to advance the interests of authors who write for the public benefit, argues that using writing as training data, in the majority of cases, is protected by the copyright doctrine of fair use, and that licensing schemes for training data are both logistically infeasible and poor policy \cite{authorguildcopyrightcomment}. Other writers have expressed distaste, but not concern, for the use of their writing in this new and unexpected way \cite{bogost2023mybookswereused}.

In this work, we focus on understanding the response of creative writers such as novelists, memoirists, poets, and fan fiction writers, all of whom engage in book-length writing that has been used as LLM training data without their consent. While focusing on creative writing may seem niche in the larger world of textual data collection, book-length writing has been found experimentally to be among the most highly valuable training data for LLMs \cite{longpre-etal-2024-pretrainers}. And, to the best of our knowledge, while there have been a series of licensing agreements between technology corporations and news publications (e.g., \cite{obrien2023openaisignsdeal}) there is no LLM available that has been trained on the consensually collected writing of creative writers. Information scholars have argued that web-scale data collection practices challenge traditional ethical principles governing the use of human subjects data, and that data should be collected and used in accordance with data subjects' perspectives \cite{vitak2016beyond}, necessitating research into the perspectives of those creating valuable data. Additionally, as the web becomes increasingly hostile to scrapers \cite{longpre2024consentcrisisrapiddecline} and content is locked down by AI companies \cite{tong2024googleredditdeal}, the need to understand what motivates---or demotivates---people to participate in consensual sharing becomes important to permitting an open web. 

With all this playing out in the public and legal sphere, we see a place for researchers to characterize the concerns of creative writers. We do not investigate these issues from a legal perspective, but rather an ethical and human-centered one, centering the voices of creative writers.
In this work, we ask:

\begin{itemize}
    \item RQ1: How do creative writers reason about the real or hypothetical use of their writing as training data?
    \item RQ2: Under what conditions, if any, would they consent to their writing being used?
\end{itemize}

To answer these questions, we conducted a grounded theory investigation by interviewing 33 creative writers. We sought to interview a wide variety of authors, in terms of genre, delivery method (i.e., method of sharing or publishing their work), degree of professionalization, demographics, and engagement with AI in their writing practice. We did not presuppose that writers should, or should not, want their writing used as training data. Instead, we sought to understand how they are reasoning about this issue and the concerns that they raise.

We found that writers reason about the use of their writing as training data according to three interconnected \corecode{principles}. 1) Writers understand that they learn from and influence each other via \subcode{the creative chain}, and question what role LLMs play in this process. 2) Writers desire \subcode{respect} for their work as time-consuming labor that requires expertise. 3) Writers note that \subcode{the human element} of writing, particularly embodiment and emotion, means that AI cannot replicate writing processes. 

These principles were often at odds with writers’ \corecode{realistic expectations} of the world they live in. 1) Writers experience a \subcode{lack of control} over how their writing is used by technology companies, and how LLMs that may be trained on their writing are used more generally. 2) Writers struggle to make sense of their contribution to LLMs given their \subcode{interpretation of scale}, when any individual’s contribution is so small. 3) Writers have both current and predicted concerns about how this technology will have \subcode{industry impacts}, some of which supersede personal impacts.

In the discussion, we consider how writing changes when it becomes data, and propose a colonial lens to understand paths forward. We also discuss the place of LLMs in the \subcode{creative chain}, and how they may (or may not) be like libraries. 
% Finally, we return to prior work on archival practices for data collection in machine learning contexts, and note how our findings point towards new principles for data collection in this context. We then address how LLMs challenge writers' notion of participating in a larger cultural project, in particular how to navigate the decontextualization that LLMs create.
Finally, we outline what kind of research might best support writing communities.

\section{Background}

In 2023, the Writer’s Guild of America, representing screenwriters, and the Screen Actors Guild, representing television and radio artists, went on strike over labor disputes with the Alliance of Motion Picture and Television Producers \cite{wgaonstrike, actorstrike}. One of the main focus points of both strikes was the use of generative AI. Although neither directly involved literary writers, it put concerns about the use of AI to displace or devalue artists into the public arena. Then, in August, 2023, The Atlantic published a story about the use of Books3, a corpus of pirated books, to train LLMs \cite{reisner2023piratedbooksrevealed}. This precipitated outrage from creative writers at the use of their work without their consent and without any compensation \cite{asmelash2023thesebooksused}. At the end of 2023, the professional organization The Author’s Guild ran a survey with 2,400 of their members, and “found nearly universal opposition among authors to their works being used to train AI systems without permission” \cite{authorsguildsurvey}.

As writers express their discontent, concerns about the legality of using published books as training data are being raised. The U.S. Copyright office started a study in 2023 regarding copyright issues raised by AI, including several requests for public comment \cite{copyrightofficeaistudy}. Dozens of lawsuits against technology corporations were filed, disputing the use of published books as training data, for the most part due to violations of copyright \cite{genaicasetracker}. Such investigations and disputes are ongoing. Legal questions will be resolved by the courts, and likely differ across jurisdictions and nations. In our work, we seek to uncover not strong legal arguments, but rather the humanistic concerns and reasoning of writers.

% In this work, we do not directly investigate the legality of using literary writing as training data, though some writers we talked to did express an opinion on the legal question. While we were open to discussing this with our interviewees, we focused our interviews on writers’ attitudes and concerns, the impact of LLMs on their career, and their perception of how LLMs might impact their industry in the future. 

\section{Related Work}

\subsection{AI Writing Assistance}

Computational processes have a long history in creative writing. Early explorations include the surrealistic Dadaism movement and the French Oulipo group, both of which engaged in algorithmic processes for writing. A 2024 review of all kinds of writing assistants found that, in the HCI and NLP research communities, papers on writing assistants have been growing dramatically since the mid-2010s \cite{lee2024designspace}. Recently, much of this work investigates how LLMs, particularly large-scale corporate LLMs like GPT-4, might contribute to particular creative writing genres, like playwriting \cite{mirowskiCowritingScreenplays2023}, storytelling \cite{ippolitoCreativeWritingAIPowered2022wordcraft}, and cartoon-captioning \cite{kariyawasam2024AppropriateIncongruityDriven}, or to particular parts of the creative writing process, like character development \cite{qinCharacterMeetSupportingCreative2024}, worldbuilding \cite{chung2024PatchviewLLMPoweredWorldbuilding}, and revision \cite{benharrakWriterDefinedAIPersonas2024}. Such work demonstrates the capabilities of LLMs to contribute to creative writing, but does little to address why such a system may or may not be adopted by writers. Nor does it address the documented negative attitude that many creative writers have towards LLMs more generally.

Another line of research investigates the social and psychological issues that AI writing assistance brings forth. For instance, research has found that increased amounts of AI generated text decreases writers’ sense of ownership \cite{leeCoAuthorDesigningHumanAI2022, dhillon2024ShapingHumanAICollaboration}.  Sense of ownership may be modulated by sense of contribution \cite{kobiella2024IfMachineGood}, the style of the AI generated text \cite{kadoma2024RoleInclusionControl}, and writers' engagement with AI contributions \cite{hoque2024HaLLMarkEffectSupporting}. Taken together, these findings almost unanimously show that, on average, AI-supported writing decreases but does not eliminate writer’s feelings of ownership, underscoring the need for a larger theory of AI participation in the creative process.

\subsection{Ethical Considerations of Data Usage}

Prior to the current interest in generative AI, researchers investigated the ethics of using social media data in HCI and the computational social sciences. Like LLM training data, social media data repurposes people’s data in ways they may not have expected and at a scale that challenges traditional principles governing the ethical use of data involving human subjects. Informed consent, for example, is often infeasible on the scale of social media datasets. Instead, information scholars argue that data should be collected and used in accordance with data subjects' perspectives on the acceptable use of their data \citep{hemphill2022comparative,fiesler2018participant,vitak2016beyond}. 

Research on social media data suggests that data subjects’ level of comfort with their data's use depends on contextual factors such as the sensitivity of the data, who the researchers are, or what the purpose of the data analysis is \citep{hemphill2022comparative,fiesler2018participant,gilbert2021measuring,gilbert2023research}. For example, \citeauthor{klassen2024black} (2022) investigate the ethics of research on Twitter data posted by Black people, finding that several participants expressed that White people doing research on specifically Black Twitter seemed colonizing, whereas the study of Black Twitter was not inherently problematic on its own. These findings suggest that writers may also be concerned about the intentions of the people collecting training data and training LLMs, as opposed to being against the use of their work as training data in all cases.

Recently, work across a variety of Computer Science subdisciplines has considered the ethical implications of existing training datasets for language models. Research has found that training datasets for popular, commercial models contains copyrighted material \cite{chang2023speak,bandy2021addressing}, contains potentially problematic content such as toxic or pornographic content \cite{gehman2020realtoxicityprompts,bandy2021addressing}, and has a skewed religious representation \cite{bandy2021addressing}. Others have investigated the representation of text from and about minoritized individuals in pretraining datasets  \cite{dodge2021documenting, gururangan2022whose,precelCanaryAICoal2024}, finding skewed representation of minoritized groups in these datasets. Researchers have argued that skewed representation in an LLM training dataset may result in disproportionate harm to minoritized groups if these models are used in high stakes contexts \cite{gururangan2022whose,precelCanaryAICoal2024}.  However, little academic work has investigated how textual data creators themselves feel and reason about the use of their writing as training data.

Just as writers are grappling with the use of their writing as training data, visual artists too are expressing concern, distaste, and often outrage at the use of visual art as training data for generative AI, specifically text-to-image models \cite{jiangAIArtIts2023}. Artists have launched a number of their own lawsuits against technology companies \cite{genaicasetracker}, often analogous to those from the writing community. Research has articulated different kinds of harm AI art is creating for artists, including direct economic harms as well as chilling effects (that is, generative AI is discouraging artists from sharing their work online, making it harder for artists to find their audience) \cite{jiangAIArtIts2023}. \citeauthor{lovato2024foregroundingartistopinionssurvey} (2024), in a survey of over 400 visual artists, found that a majority of artists believe training data for generative models should be disclosed and express concerns about the impact of AI on their industry \cite{lovato2024foregroundingartistopinionssurvey}. We see this work as a primer for our own, as creative writers may or may not reason about generative AI in the same way as visual artists, and individual and industry impacts may differ.

\subsection{Addressing Concerns about Data Usage}

Technical research attempts to address artists' concerns about training data. To address concerns that generative AI may plagiarize pretraining data contributors, researchers have proposed approaches to minimize model memorization \cite{ippolito-etal-2023-preventing,carlini2023quantifying,kandpal2022deduplicating,lee2021deduplicating,vyas2023provable,chu2024protect} and style mimicry \cite{shan2023glaze,liang2023adversarial,van2023anti}. Another line of work attempts to protect the IP of writers or programmers by watermarking text \cite{lauprotecting, qiang2023natural, yoo-etal-2023-robust} and attributing generated content to specific training data \cite{dengcomputational,hammoudeh2024training}. To enable individuals to opt out of training data, researchers propose methods for unlearning \cite{chen-yang-2023-unlearn,jang-etal-2023-knowledge} and modular language model architectures \cite{gururangan-etal-2022-demix,li2022branch, minSILOLanguageModels2023}. Such work presents potential technical solutions, but does not answer questions about whose work should be protected and under what contexts. 

Several researchers have applied learnings from the library sciences about how to collect and archive data to machine learning contexts. \citeauthor{jo2020lessons} (2020) outline five important aspects of data collection from archival and library sciences, and how they may be applied in the machine learning setting, such as actively collecting underrepresented data and allowing participants to denote sensitivity and access levels for their data. Relatedly, \citeauthor{desai2024archival} (2024) take an archival perspective on pretraining data explicitly, arguing that “to the extent that people use LLMs as interfaces into history and culture, the selection of data shapes and constrains that experience” just as archival collections do. While some advocate for more attention and care to be put into data collection \cite{gebru2021datasheets}, it has been documented that incentivizing data work is difficult in research and industry communities \cite{sambasivan2021everyone,gero2023incentive}.

% To the extent that a subset of the Computer Science community cares about data work, for instance documenting data provenance \cite{longpre2024large}, popular corporate LLMs give few details about the data used to train their models \cite{}. Such opacity makes it difficult for writers to know for sure whether or not their work has been used, and prevents avenues for recourse.

\section{Methodology}

We engaged writers through hour-long interviews which were primarily conducted virtually. One of the authors conducted about two thirds of the interviews; another conducted the other third of interviews. Interviews were automatically transcribed, and then listened to by one author in order to clean up errors in the transcript. This meant that at least two authors listened, in full, to each interview (the original conductor of the interview, and the person who cleaned the transcript). Additionally, interview transcripts were often read in whole by a third author. The first four authors participated in the coding, and all authors participated in discussions about the codes.

\subsection{Procedure and Analysis}

We employed Grounded Theory to conduct the interviews and analyze the data. Because the details of how Grounded Theory is employed can vary depending on school of thought \cite{mullerGroundedTheoryMethod}, rather than describing our approach as belonging to a certain lineage we instead detail our procedure. We began coding and theorizing about the interviews shortly after we began conducting them, which allowed us to constantly compare our theories with incoming interviews and update our interview guideline as the study progressed. Our initial interview guideline was based on concerns we had heard writers bring up in a variety of domains: as reported in the news, as noted in ongoing lawsuits, as represented by professional organizations such as the Author’s Guild and the Author’s Alliance, and as mentioned to us by writers we knew in our professional and personal networks. Although traditional Grounded Theory eschews existing literature, our research intended to directly investigate a growing issue. Therefore, we could not approach the data with “an empty head.” Instead, we approached it with “an open mind” and indeed found that writers' reasoning and concerns about these issues did not always match the way they were represented by the news, lawsuits, or professional organizations.

The interview guideline was, thus, updated as emergent themes developed through coding. For instance, our initial guideline did not include questions about the nature of the institution collecting the training data (for profit, non-profit, research, etc.). However, early on in the interviews it was clear that some writers cared a great deal about the type of institution, and so we updated our guideline to explicitly ask about this in subsequent interviews. Ultimately our guideline was updated five times, and the final guideline can be found in \autoref{app:interviewguideline}. 

Initially we employed open coding, but found that our interviews covered a lot of ground and open coding was, in fact, too open. Drawing on the approach of \citeauthor{deterding2021flexible} (2021), we then performed “index coding”, where we coded interviews based on the themes that were explicitly asked about in the interview guideline. Our developing interview guideline acted as the first set of codes. Then, individual index codes—such as attitude towards compensation, or concerns about reproduction of writing style—were collated and open coding was done within a single index code. This more narrow open coding allowed us to deeply investigate specific writerly concerns.

Then, axial and selective coding was done on all codes that came out of the open coding performed on each index code. This allowed us to see emergent and general themes that were occurring across the  concerns that writers were bringing up. With these new, general codes, we performed a final round of coding in which all excerpts were re-coded with the general codes to both test their performance on the data, and to note any data that did not fit into this framework. These codes are what are reported as our results.

\subsection{Recruitment}

Writers were recruited in a variety of ways: through our professional and personal networks, through cold emailing writers from specific categories (e.g., romance novelists), through posting on writer forums (e.g., forums for writers that use AI in their writing practice), and through snowball sampling where we asked interviewees to suggest other writers to interview. All writers were from North America.

We initially wanted to recruit writers from a variety of genres (fiction, nonfiction, poetry, and fan fiction) and who publish in a variety of methods (big 5 publishing, independent publishing, self-publishing for profit, and self-publishing for free). Throughout the study we added three more types of variation we wanted to seek out: level of professionalization (majority of income comes from creative writing, majority of income comes from other writing-intensive profession, majority of income does not come from writing), self-described demographic information (race, gender, sexuality, and religion), and engagement with AI in writing practice (none, has tried it out, regular use in creative composition, and regular use for paratext e.g., marketing materials). Therefore, we engaged in theoretical sampling where we recruited participants based partially on what was coming up during our analysis. Note that we did not explicitly ask for demographic information (e.g., sexuality); instead a writer might’ve brought up self-described demographic information as being relevant to our questions, for instance noting a lack of writers like them being represented in training data. In \autoref{app:demographics} we give detailed accounts of the distribution of writers.

Although we did not recruit for this specifically, we found that our participants had a wide range of views on language models more generally. Some considered the technology to have no positive applications, others considered it to be neutral, and still others thought it had pro-social applications, even if they were against the current way in which it was created or used. 

% Finally, although we purposefully recruited a diverse sample of writers, our intention was not to investigate or make claims about how certain perspectives may correlate to different profile attributes, nor does our methodology support making such claims. In our findings, we provide details about participants where it is necessary for understanding their reasoning.

\subsection{Ethics Board Approval, Privacy, and Anonymity}

This study was approved by the relevant ethics review board. Interviewees were anonymized, both by the removal of their names as associated with coded excerpts, as well as by removing or altering any details that may be identifying, for instance a description of their published work. Finally, although video and audio were recorded, only anonymized transcripts were retained, with video and audio recordings deleted within one year of the completion of the study.

\subsection{Positionality Statement}

All of the authors of this paper are based in North America, and all of our interviewees were similarly based in North America—either the USA or Canada. Although training data for LLMs can come from across the world, there is some evidence that most of it is written by those in North America \cite{gururangan-etal-2022-demix, dodge2021documenting}. However, we note that the cultural context of North American writers may be different to those from other parts of the world.

The research team contained researchers from Computer Science, Information Science, Sociology, Rhetoric and Composition, and Law. In addition, several members of the research team have their own creative writing practice.

\section{Findings}

All our codes can be found in \autoref{tab:findings}. First, we report on the \corecode{principles} that are the foundation for how writers think about writing in general, and that they use to reason about the use of their writing in this new context. Second, we report on their \corecode{realistic expectations} about how the world works. Here, we see writers’ principles can sometimes, but not always, clash against what they think is possible in the world they live in. 
% Third, we report on \corecode{idealistic futures}, in which writers attempt to thread the needle of their principles through their realistic expectations. 
% These futures represent an attempt to align their principles with realistic expectations about the world, and envision ways that writing may be used as training data in ways that are positive for all stakeholders.

% \corecode{Principles} and \corecode{realistic expectations} are broken down into a set of codes that emerged from our analysis. However, w
We note that there are a variety of ideas that thread throughout multiple codes. For instance, a major discussion point was if and how compensation may work. This idea comes up under the code \subcode{respect} (where compensation is often seen as a mechanism of showing respect) as well as under the code \subcode{interpretation of scale} (where writers reason about how much compensation could reasonably be expected) and \subcode{lack of control} (where writers note that it may be impossible to control how their writing is used, negating the possibility for fair compensation). Our codes outline how writers are reasoning about these issues, rather than presenting a list of concerns.

\begin{table*}[h] 
    \centering 
    \caption{Overview of the codes that emerged from our grounded theory analysis.}
    \label{tab:findings}

    \begin{tabular}{l p{11cm}} 
        \toprule
        \textbf{Principles} & \textbf{} \\ 
        & \\ \

        The Creative Chain & “\textit{It's just \underline{being a link in the creative chain}. I made something. Somebody else saw that, and they got a spark of an idea to do something else, and they want to build off of it. We're all building off of each other constantly.}” (P20)\\ 
        & \\ \
        Respect & "\textit{I think it just comes again to \underline{respecting the works and respecting the people}. Because I think that's probably my main issue with a lot of AI stuff is just the complete lack of respect for the works that they have been using to profit off.}" (P18)\\ 
        & \\ \
        The Human Element & "\textit{Honestly, I think a lot of the reason that people like reading other people's stuff is \underline{because there's the human element there}. You're looking at one guy's perspective of an insane situation ... It's the project of one person's mind.}” (P13)\\ 
        & \\ 
        \midrule
        \textbf{Realistic Expectations} & \textbf{} \\ 
        & \\ \

        Lack of Control & "\textit{One of the things that we're all really feeling right now is \underline{a lack of any options in this}. When I found out that my work was in Books3 I emailed my agent right away. And I was like, I know there's like nothing you can do, I just want you to know that this is a thing, and that I don't like it.}" (P20)\\ 
        & \\ \
        Industry Impacts & "\textit{We could easily be putting artists out of work. And what does that mean for them as artists and humans, [who need] to put food on their table, but also like, what does that mean for culture and society?}" (P9) \\ 
        & \\ \
        Interpretation of Scale & "\textit{Cause I'm trying to think, okay, upsides and downsides. It's hard to pinpoint one, because again, \underline{I'm a little drop of water in a huge ocean}. I don't have a distinct style, so I don't... I don't really know.}" (P24)\\ 
        & \\ 
        \bottomrule 
    \end{tabular}
\end{table*}

\subsection{Principles}

\subsubsection{\subcode{The Creative Chain}}
\label{sec:creativechaing}

Participants often compared making their writing accessible to train a language model to their principles and expectations around making their writing accessible in other contexts.  In general, participants recognize that reading, imitating, referencing and transforming other writing is essential both to learning to write and writing itself. In turn, many participants described a responsibility to support education and the production of art by making their writing accessible and taking a permissive stance towards work that references theirs. As P12 (professional novelist) put it, “\textit{I think that being very rigid around intellectual property, and the defense of ideas…  turns into an awful scenario that I don't really believe in, because I believe that all art inspires other art, and I love that basic premise.}” P20 (professional novelist) described this as “\textit{being a link in the creative chain}”: “\textit{We're all building off of each other constantly, you know. My work is just the product of everything I've absorbed through my life… nothing makes me happier than seeing people make other things with my work.}” 

Making information free and accessible for others to learn from was an essential principle for several participants. Because of this, some writers felt ambivalent or positive about their writing being used to train language models, despite other concerns they might have (P5, P17, P24, P25,P27, P30). 

\supersubsection{Expectations of control in the creative chain: What is the project I am a part of?}
In contrast to other creative projects a writer might contribute their work to, some participants expressed that the overall goal of language models was unclear to them, and it was therefore difficult to assess what it meant to contribute their writing. P3 (poet and essayist) noted that “\textit{it feels similar to what I imagine an artist who's being curated might feel: … they would want to know about the content or subject, or the nature of the thing that they're being curated into.}" Understanding the goal of the project was important for P14 (poet), who said, “\textit{If I knew the end result or kind of, Here's why we're interested in your stuff,  I would be much more interested in opting in.}”

\supersubsection{Disagreements about the artistic potential of language models.} 
% P15: if I imagine a version of this that is interesting and provocative, and I know all these	artists that have made interesting work from it. I think I'd be more inclined to be like well, I'll be part of that weird thing.
Participants disagreed about whether contributing to training datasets could yield benefits such as supporting the production of worthwhile art, helping others to express themselves, or supporting education.  Several participants had trouble imagining the artistic potential of language models because they did not like the style of generated text (P21, P25), found it “\textit{generic}” (P6, P13,  P24), or had concerns about the political biases of these models (P18, P21). Many others expressed doubts about the creative potential of LLMs, both because they view LLMs as “\textit{regurgitation}” (P18, fan fiction writer) of other work rather than creating new ideas, and because of the fundamental role of humanity in creativity. (See \subcode{the human element} \ref{sec:humanelement}.)

On the other hand, several writers reported regularly using LLMs in their creative process (P1, P8, P15, P28), or cited literary work that used language models that they liked (P4, P12, P15, P22). Many others saw the value in using LLMs “\textit{as a tool}” (P2) for creative expression, imagining that LLMs might help writers generate initial ideas (P18, P27), provide writers with feedback or analytic summaries during the writing process (P3, P5, P16, P27), or help writers move past writer’s block (P3,  P13, P15, P18,  P29). 

\supersubsection{Representation and nourishing the future of technology.}
In addition to supporting the production of art, writers identified other positive, meaningful aspects of contributing their writing to training data. Some writers expressed excitement about the idea of “\textit{contribut[ing] to the future}” (P28, poet) or “\textit{nourish[ing]}” (P15, novelist and journalist) language models with literary writing (also P1, P26). As P11, a professional novelist who had never used an LLM,  explained, “\textit{I've had an influence on the way this machine thinks, which is incredible. I mean, it's probably that's my place in history more than any encyclopedia entry, you know.}” 

Some writers with racial identities that are under-represented both in language model outputs and in the literary world more broadly were enthusiastic about contributing their data (P23, P27), as “\textit{there needs to be more diversity in some of these models}” (P27). 

\supersubsection{Decontextualization: being included could misrepresent or reify a writer's style, work, and identity.}
Some participants expressed concerns about how contributing to language models had the potential to decontextualize their work. For P24 (novelist), decontextualization is a threat to their artistic evolution and identity: “\textit{I don't want my past books to be training… because I want to always evolve and become a better writer, and I feel like that would leave me stuck in that same voice or ability}.” P32 (professional essayist and novelist) also expressed concerns about decontextualization, as they thought through what would come up if someone prompted a model trained on writing they wrote throughout their gender transitions. They were comfortable with someone reading their old work online, where it would be accompanied by contextual information like when it was written ("\textit{I trust readers, I think readers are really smart}"); in an LLM, they feared it could instead be weaponized against their community.

For P19, the way language models racially decontextualize their training data was a big concern: 

\begin{quote}
    \textit{Specifically as a Black writer, Black scholar, Black cultural worker and artist, if these models are being trained to  give an output and the way that the ingesting happens is  taking from a lot of different places, I think it supports this idea that ... we can sort of move towards a  universal---either multiracial or deracialized---art. And then that, of course, is going to support people in positions of power, specifically  white people.} (P19)
\end{quote}

\supersubsection{Expectations of respect and humanity in the creative chain.}
Participants made analogies between contributing their writing to language model training datasets and various ways their work does or could relate to other creative projects: it is like someone turning your book into a movie or play (P3, P12, P20), it is like having your style mimicked by your student or competition (P6, P7, P31), it is like having your words appear uncredited in someone else’s work (P12, P13, P17, P18, P29, P31), it is like someone writing fanfiction about your novel (P1, P18, P20), it is like being curated into someone’s art show (P3).  Participants described how these different scenarios present different risks to their careers, reputations, and sense of creative control. As such, in each of these scenarios, participants described how they expect to receive respect through some combination of credit, compensation, creative control, or at least the sense that a human being who admires their writing engaged with it, expressing a shared value of creative writing.

\subsubsection{\subcode{Respect}}
\label{sec:respect}

% \begin{quote}
%     \textit{I think just comes again to respecting the works and respecting the people. Because I think that's probably my main issue with a lot of AI stuff is just the complete lack of respect for the works that they have been using to profit off.} (P18)
% \end{quote}

While writers understood that by publishing their work they are allowing others to use that work in perhaps unexpected ways, they also felt that the use of their work as training data indicated a lack of respect both for writing as an endeavor and writers as people. Although writers did not express that all uses of their writing require consent---for instance, as in the section above about the creative chain, they expect that others will learn from their work or perhaps even try to mimic it---the use of their writing as training data felt different. P6, a professional romance novelist, said that the use of their writing for AI may be more like signing away their foreign subrights, which would require consent and be part of contract negotiations when publishing. 

\supersubsection{Mechanisms of respect: compensation is key for professionalized writers.} The idea that writers wanted respect was very much tied to the mechanisms in which respect can be shown. 
% These mechanisms included consent, compensation, credit, and involvement. 
The mechanisms that writers desired was a function of many things: their degree of professionalization, the type of writing they worked on, their sense of precarity in the labor market, and their perception of the organizations that were or could be using their writing.
Primarily, writers wanted respect for writing as a laborious act on its own---to write a book takes time, effort, and expertise. As P9 (professional romance novelist) put it,

\begin{quote}
    \textit{The biggest myth about creators, artists, writers, you know, name the person who creates art or entertainment in the world, is that we do it for the muse. Like, I live in this house, and I have to pay a bill every month to pay for this house, you know, I have to put food on my table and clothe my children and myself. Like this is my job... We should pay people for their work, right?} (P9)
\end{quote}

The people we talked to who were professional writers---who earned the majority of the income from their creative writing practice---were most likely to cite compensation as the main mechanism of respect they would expect. For instance, P8, a professional freelance writer, noted that, “\textit{And I say this as a, you know, a very pro AI person who thinks it's the next creative way. But content producers actually need to be compensated on some level.}” P11 (professional journalist and novelist) made the comparison with other kinds of expertise: “\textit{If NASA wanted my expertise to hire me to help in the space program they would actually have to pay me quite a lot of money, and they would do so.}” However, technology companies were not treating writers as if they brought any kind of expertise or as if their labor was valuable.

\supersubsection{Power differentials modulate desire for compensation.} Writers were also quick to point out the large power differential between the corporations creating these models and the writers trying to earn a living. For instance, P7 (professional journalist) explains that, “\textit{The tech companies are making money hand over fist, and I feel like they can compensate people.}” Professional writers were likely to say they would require compensation regardless of whether or not the model creator was making money, but many writers did make a distinction between organizations they perceived could afford to compensate writers, and non-profit organizations or applications. We return to the idea of who can afford to compensate writers in the realistic expectations section.

\supersubsection{Different kinds of writing require different kinds of respect.} Writers discussed genre distinctions in terms of the level of labor involved. P9, a professional romance novelist, talked about the use of certain kinds of books feeling particularly disturbing, “\textit{like Pulitzer Prize-winning work that's happening in nonfiction},” which represents decades of research then being taken without consent or compensation and made “\textit{part of the churn}.” In this case, writers wanted respect for time-consuming labor that writing represents.

Other writers noted that the use of certain kinds of writing felt more disrespectful than others because of the contents of the writing, rather than the labor taken to produce it. P4 (fan fiction writer and poet) talked about some writing being “\textit{deeply personal}” and being “\textit{disturbed by the lack of care}” that is shown by the act of putting such work into a language model. P32, in discussing a colleague who had written a nonfiction book which included accounts of assault, noted that their feelings and words about the person who assaulted them now lives on in the predictive models trained on that book, saying, "\textit{There is an aspect of that that is like literal haunting}." Here, respect was not about compensation but about understanding the sensitive nature of some kinds of writing.

\supersubsection{How to value writing.} Writers discussed how they might consider whether or not to contribute certain works of theirs, which led them to express how different kinds of writing may be valued differently. P14 (poet and translator) said that if they were asked for consent, “\textit{there are things that I would probably hold back in order to wait until I've gotten the value out of them and then at that point I wouldn't worry about it as much}.” Similarly P13 (fan fiction and science fiction writer) noted that they used to post a lot more of their ideas on social media before they started winning awards for their stories; now, they consider their ideas to be more valuable and are more hesitant about sharing them publicly. Other writers noted that older writing is often less valuable, either, as P14 noted, because they have already gotten value out of it, or because older writing is worse---they have become a better writer and therefore value their more recent writing more (P7, journalist and novelist).

The desire for respect was very much about respecting the humanity of the person who created the work. (As P32 noted, everyone deserves respect.) Implicitly, writers were often complaining that the use of writing as training data without consent or compensation indicated a lack of recognition that people had created the writing. When text can be scraped from the web, it is easy to forget how that text was created, or to recognize the text as more than “just” data. This led writers to articulate what they consider special about their writing compared to other kinds of writing, including text that may be generated by computers, which we report on next.

\subsubsection{\subcode{The Human Element}}
\label{sec:humanelement}

Large language models generate text because they are designed to respond efficiently and accurately to a prompt via token prediction. If a human writer gave this as a reason for why they write, they would be, as Shannon Vallor notes, “doing it wrong” \cite{vallor2024dangernotwhatyouthink}. The writers we spoke with saw their writing as fundamentally different from the writing produced by LLMs, because of their humanity and readerships that similarly value and respect their humanity through their writing. With this principle, we aim to roughly capture how and why writers distinguish their writing from the text generated by LLMs. Creative writers have many reasons for why they write, but notably, none of them are about efficiency or prediction. Instead, some of the many reasons cited include: creating art (P15, P18, P25), making a living (P8, P9), processing and sharing emotion (P10, P16), expressing one’s experiential, embodied subjectivity (P5, P19, P27), and improving one’s thinking on and understanding of culture, place, society, and oneself (P13, P14, P29). Underlying each of these reasons are three key sub-principles: 1) that writing is a fundamentally human activity distinct from the text generation of LLMs, 2) that the humanity of writing is valuable to both writers and readers, and 3) that the ‘readership’ for a training dataset is not the same human audience that writers wish to reach.

\supersubsection{Writing and text-generation are not the same.} The parsing and production of text is often enough to garner comparisons between creative writing and text generated by LLMs. However, many writers point out that this is a false equivalence. Motivations and methods for each differ dramatically. 

Many writers see the text generation of LLMs as an attempt to mimic human writing, and, in many cases, an author’s unique style or voice. This is tied to discussions of \subcode{the creative chain} (\autoref{sec:creativechaing}) and writers view the training of LLMs on a particular writer’s voice as impoverished and decontextualized. When they read writing that inspires them, writers appreciate, contextualize, question, compare, and transform a text within their own social, embodied, and emotional contexts, something mass aggregates of textual data cannot do. P6 (professional
/novelist) contrasts these processes: “\textit{My absolute favorite book was instrumental in the way that I wrote the favorite of all of the books I've written. But it's still coming through me. Whereas, I feel like ChatGPT would end up poorly mimicking me, as opposed to being something that came from someone else, but was formed by their appreciation for me}.”

Several writers express that what LLMs do when they generate text in an existing author’s voice is like a “cheat code” for writing and that a person’s experience and perspective enable real writing to happen. P17 says: 

\begin{quote}
    \textit{Write your own goddamn book. Like, you gotta go through it. That said, my writing's very personal, it's very autobiographical. So, no one's gonna write my book, I get that. But it does feel kinda like a cheat code, you know? Like, I shouldn't be able to go in and [ask a model], ‘Hey, rewrite this scene in the voice of Cormac Mccarthy.’ Cormac spent his whole life doing that, so why should I be able to just rip it off?} (P17)
\end{quote}

Implicit in these distinctions is an understanding that LLMs do not have authorial intent, which is central to writing as a human activity. That LLMs do not have authorial intent when generating text is viewed as an asset to those writers choosing to use them in creative ways. P31, who has used LLMs in their creative practice, notes this explicitly, while problematizing how LLMs are colloquially talked about as “collaborators” in some contexts. They say: 

\begin{quote}
    \textit{I'm troubled by the idea that it's a collaboration or a co-writer. I don't think like that when I'm doing it. That’s attributing maybe a bit too much power or agency to it. I don't ascribe [a] pseudo sentience to it or anything like that. … There's no will on the other side to make something.} (P31)
\end{quote}

This is a positive thing for all of the writers we spoke with who use LLMs creatively. Because the writer has human agency and the LLM does not, writers maintain both control and uphold the value of humanity in the creative process.

\supersubsection{Writing’s humanity makes it valuable.} Almost all of the writers we spoke with expressed some variation on the idea that at least part of the value of writing comes from the fact that it was written by a person, to be read by another person. Writers consistently offered things that readers value in writing---a unique style, voice, or perspective, authenticity, and humanity chief among them. What writers think readers of creative writing value is not easily replicable by LLMs. “\textit{I write poetry as a reason to connect with other people,}” P16 says. P13 remarks: “\textit{Honestly, I think a lot of the reason that people like reading other people's stuff is because there's the human element there. You're looking at one guy's perspective of an insane situation ... It's the project of one person's mind.}”

Writers who use LLMs as part of their creative processes also see distinct differences between the motivations and methods of writers and LLMs. Some of these writers use the ‘mimicry’ that LLMs enact as valuable parts of their process. P25 describes how they use the mimicry of LLMs in their work positively, after inputting their writing as source material into an LLM: 

\begin{quote}
    \textit{
    % I want the machine to mimic the way that I'm daydreaming about whatever piece [I am working on]. Then 
    I want the machine to start telling me things that I haven't [thought about], like a little thing for me to come up with a better question. Once I come up with a better question, not for the machine but for myself, then I either go and look it up or I'm like, ‘Great, now I'm gonna go think about it.’} (P25)
    % and then I go and work.} (P25)
\end{quote}

\supersubsection{The audience for a training dataset of creative work is not the audience that writers want or imagine.} Undergirding our conversations with writers was the understanding that the imagined or intended audience for a writer’s work is other people, not a dataset using writing to train an LLM. Every direct mention of the words ‘audience’ or ‘reader’ in our interviews was in reference to an explicitly human audience. At least one of these two terms was mentioned in all of the interviews we conducted. 

In light of the strong connection between humanity and writers’ conceptions of audience, writers were left to conjecture about who the ‘audience’ of training data really was. Algorithms, corporations, and users of LLMs were all variously floated as ideas of what might constitute the readership for training datasets. P26 wants to share their work with people, but is “\textit{not ready to put [even] a little piece of it into the algorithm.}” The scale and combinatory nature of training datasets provoked skepticism and sometimes fear around this imagined ‘readership,’ quite unlike the readerships writers imagine for themselves.

\subsection{Realistic Expectations}

\subsubsection{\subcode{Lack of Control}}

\begin{quote}
    \textit{One of the things that we're all really feeling right now is a lack of any options in this. When I found out that my work was in Books3 I emailed my agent right away. And I was like, I know there's like nothing you can do, I just want you to know that this is a thing, and that I don't like it. And if there are any conversations on the publisher side, on the agency side, anything where I can add my voice to that, anything we can take action with, I want to do that. Then he wrote back and he said, I know, I've had this same email dozens of times this week. Everyone hates this, nobody wants their work being used like this, but we have no way to opt out. There is no legal recourse for it.} (P20)
\end{quote}

All writers emphasized the lack of control they had in this situation. Lack of control over what is done with their writing more generally, as well as lack of influence over or trust in how technology corporations might make use of their work. While writers acknowledged that the use of their writing in unexpected ways is, in fact, an expected part of publishing---i.e., being part of the creative chain---they also expressed that certain use cases are regulated. 

\supersubsection{Certain uses of writing are restricted and there are avenues for recourse.} Writers often sign contracts that outline a publisher's right to republish a book in a foreign country, or create a movie adaption, indicating that writers do have control over a variety of use cases of their writing. For instance, P11 (a professional novelist and journalist) said, “\textit{There are lots of sources of pirated books on the Internet, but they're illegal, and we try to shut them down. If ever I find a free copy of my book circulated somewhere, I'll ask my publisher to send a cease and desist letter, and it usually gets taken down.}” %In addition, as P11 pointed out, there also exists recourse when such rights are not respected, as in taking down online copies of pirated books. 

Currently, the use of writing as training data for LLMs is both outside of writers’ control and they lack any recourse when their writing is used in this way against their wishes. Writers understood that this lack of recourse had several sources. In addition to their lack of control, writers agreed that, even if such use was restricted, it is difficult, if not impossible, to stop models from being trained or remove models that already exist (P2, P6, P12, P32).
% As P6 expressed, “\textit{I don't see it getting pulled back. And I think the day is going to come when it's harder for me to make a living.}” 

\supersubsection{It is difficult to understand how to restrict only certain use cases of LLMs.} Writers did not always disagree with the creation of language models per se, and instead had concerns only about certain use cases. But many writers noted that it seems impossible to prevent certain use cases. For instance, P24 (a fantasy novelist) would consider consenting to the use of their writing if they could ensure that the model would not be used to compete with them, but felt “\textit{like that would be an impossible ask. People are just gonna lie and go ‘No, I'm not writing a novel.’}” Others noted that while it may be possible to audit big companies, “\textit{there are all these people putting out small models, anyone can do this now, and it almost feels impossible to rein in}” (P2, a writer experienced with LLMs). 

Several writers said that they would be happy for their writing to contribute to prosocial models, for instance models that supported literacy efforts or were specifically designed to help people without access to other forms of education. Yet, they noted that they have no control over who makes the models or what they are used for. In this way, they called upon the idea that they were not being involved in model making, which is one of the ways writers wished to be given respect for their work and their expertise. 

\supersubsection{Distrust in technology companies furthers the belief that restriction is impossible.} Others noted that they didn’t trust that even if they were promised that certain use cases could and would be restricted, that those promises would be upheld. “\textit{The way that it's been working so far is that people take stuff, and then later we'll be like, Oh, my God, I'm so sorry! It's too fucking late, it already happened}” (P13). One writer articulated how their understanding of the actions of technology companies in the past influences their expectations with this particular issue:

\begin{quote}
    \textit{This is what tech companies have always done. Uber is like, we're just gonna make these cars. And we'll see if cities can catch up with us. Or Google Maps when they're like, we're just gonna photograph everybody's house. And then we'll see if anyone gets mad. I think that's totally what AI writing software is doing. They're waiting to see what people are gonna do.} (P12)
\end{quote}

Finally, writers acknowledged the difficulty of predicting future use cases. As technology changes, it can be hard to know what you might be agreeing to in the moment. Some established professional novelists were especially concerned about those at the start of their careers---established writers were concerned about up-and-coming authors (P9, P12, P20) and writers we interviewed who were at the start of their careers noted their particular predicament (P13). As an early stage writer, one has less power to negotiate certain terms out of their contract and may end up agreeing to terms that, years down the line, perhaps when the writer is more famous, are unexpected, unprecedented, and ultimately undesirable.

\subsubsection{\subcode{Industry Impacts}}

% \supersubsection{Present day impacts of language models on the creative writing industry.}  
Though public discourse about the impact of language models on the creative writing industry often takes place in the future, writers described present day impacts of language models on their careers and fields. P12 described the strain language models have placed on literary editors, who “\textit{spend hours every week}” sifting through large amounts of “\textit{junk pitches and generated stories.}” P12 recalled how this influx of generated work led one prominent literary magazine (Clarkesworld) to “\textit{close submissions entirely,}” which “\textit{really hurts people who are just getting started, people who already have this narrowing slate of options where they can submit their work.}” Other writers expressed similar concern over the impacts of language models on new writers getting started in the industry (P9).

Writers with academic teaching positions described confronting the impacts of language models on creative writing education. These writers explained how their creative writing careers and teaching careers are inextricable, either because they are hired to teach because of their creative writing careers, or because they see teaching as part of their creative practice. These writers note that even if they are morally against LLMs, for instance because of the energy used to train and use them, they must deal with this technology in their classrooms (P12, P14, P16, P18, P32).

Some writers also described positive impacts of language models and AI on their careers. Self-published authors described using language models and other forms of generative AI to help with the work a traditional publisher might do, like creating a cover or producing marketing materials for their work (P24, P27). P31, who makes use of language models in their work, described how their career has benefited from advances in LLMs and the resulting increase in public interest in AI.

% \todo{contract stuff: P20 I have heard through the grapevine, too, about debut, authors, new authors, people who are getting booked contracts for the first time. Finding language in their in their contracts about permission to use these works on for for training these models}

\supersubsection{Predicted future impacts.}
By and large, most participants found that in their current state, language models are not direct competition for writers as they do not produce sufficiently sophisticated, high quality writing. When speculating about how future advanced versions of language models might impact their careers and industries, many writers imagined competing with a language model that had been trained to mimic their writing. Other writers speculated more broadly about how language models might compete with or displace writers, i.e., "\textit{an AI that writes better novels than humans}" (P12, professional novelist).

Writers expressed different perspectives about whether or not these potential futures were a threat to the creative writing industry in general and their careers in particular. P6, a romance novelist who makes her living by selling her novels, finds these futures imminent and threatening to her livelihood: “\textit{I asked [my finance guy] how fast we could move up my retirement...I think the day is going to come when it's harder for me to make a living, because so many people can do exactly what I do}.” Other commercially successful writers speculated that because their readers are interested in their specific style or voice, they “\textit{have very little to worry about}” (P9). As P11 put it, “\textit{I really don't see why anyone would look to buy an AI generated book by [P11] when mine are readily available.}” 

Several authors expressed a lack of concern about the commercial or monetary impact of LLMs on creative writing, because of the value readers place on the humanity behind a creative work. P10 emphasizes the elisions and false equivalencies drawn between writing and text generation in their response, underscoring the value of writing as a distinctly human enterprise. 

\begin{quote}
    \textit{I don't think [LLMs are] particularly useful for the creative industry. I mean, I could see them being useful for, like, ... discovering a new kind of a new antibiotic or something like that... But, human expression? I guess I think of poetry or fiction as a deeply human activity. And it sort of baffles me why people would want to read what has been historically a human activity that has been produced by a machine.} (P10, poet and essayist)
\end{quote}

Because of the importance of a human perspective in creative writing, some writers do not believe that language models could ever “\textit{evolve to where … [it can] come up with new stories that are relevant to the current readership}” (P24). P11 imagined that while people would be interested in, for example, seeing the first movie with an AI-generated script, the novelty would soon wear off, and LLMs would just become “\textit{part of our toolkit to make good art}”.
% ; Several writers imagined that rather than competing directly with LLMs, LLMs might become “\textit{part of our toolkit to make good art}” (P11). 

\supersubsection{Backdrop of a changing industry.}
Some writers acknowledged that their resistance to contribute to training datasets was “\textit{compounded by a sense of precarity}” in the creative writing industry (P3, poet). P20 (professional science fiction novelist) noted that “\textit{it's a really difficult time for people in our profession and for creative industries in general. The job market is constantly shrinking.}”

Some writers expressed protectiveness towards the creative industry, citing the value of humanity intrinsic to creative work. The questions these writers were asking were existential, implying that the existence of AI-generated art and literature is forcing us to ask, ‘What kind of world do we want to live in?’ P9 (professional romance novelist) notes: 

\begin{quote}
    \textit{We could easily be putting artists out of work. And what does that mean for them as artists and humans, [who need] to put food on their table, but also like, what does that mean for culture and society? The argument from the AI side is, ‘Well, this is culture and society too. This is just advancement.’ But, as an artist, I'm always going to say humans make art because it's about feelings.} (P9)
\end{quote}

P13, along with several other writers we spoke with, wants to live in a world where people can make a living as writers: “\textit{I don't think it's good to make it cheaper to make art for the fact that people need to be employed and I would prefer to be employed as a writer than as a lawyer.}”

Others expressed that since they did not expect commercial success as a creative writer, they felt “\textit{ambivalent}” (P14) about the impacts of language models on their careers. As P14 put it, “\textit{You know, I've made \$20 on poetry over the course of my life. So. It's not, yeah, I'm not on [the commercial] side of [writing] or interested in that.}” P30 similarly agreed that for poets, “\textit{the stakes are different}”: “\textit{as a poetry MFA student … we're not doing it for the career prospects}.” 
% \todo{something here that’s like… these people are not worried about losing the humanity of writing; when asked how P30 felt about someone using a language model to write a poem he said “I’m glad people are writing poems”, connection to creative chain; then some kind of transition to next section}

\subsubsection{\subcode{Interpretation of Scale}}

% \begin{quote}
%     \textit{Cause I'm trying to think, okay, upsides and downsides. It's hard to pinpoint one, because again, I'm a little drop of water in a huge ocean. I don't have a distinct style, so I don't... I don't really know.} (P24)
% \end{quote}

The amount of training data used to train large language models is hard to comprehend, and often, in fact, opaque to those outside of the corporation that trained the model. To briefly contextualize this section, we point out that GPT-3 was trained on ~700GB of text \cite{brownLanguageModelsAre2020}, and newer models are trained on significantly more, such as Llama 3 which was trained on over 15TB of data \cite{dubey2024llama3herdmodels}.
% a commonly used, open-source training datasets for language models is The Pile, which is about 800GB of text \cite{gao2021thepile}. 
% Researchers have investigated the estimated amount of data needed for models of certain sizes, reporting that GPT-3 was trained on approximately 300Billion tokens, or an estimated 1TB of text (assuming 1 tokens ~= 3 characters) [Training Compute-Optimal Large Language Models]. 
% Another relevant reference is Books3, a corpus of pirated books which several of our interviewees reported included their books. Books3 has been used, in collation with other datasets, to train a number of LLMs [cite]. Books3 contains almost 200,000 books or 200GB of data \cite{biderman2022piledatasheet}. 
If we assume the average book contains about 100,000 words, then a single book would be the equivalent of 0.1MB, or ~0.0000001\% of the training data of GPT-3. Such magnitude is difficult to reason about; as P24 said, "\textit{Cause I'm trying to think, okay, upsides and downsides. It's hard to pinpoint one, because again, I'm a little drop of water in a huge ocean.}"

\supersubsection{Compensation is hard to determine and likely to be insignificant given the scale of LLMs.} 
While most writers objected to their data being collected by corporations, they simultaneously had difficulty comprehending how to  value their contribution when considering the scale of training data used. Some noted that their style was not distinct enough to be noticeable in any outputs or to significantly change the model (P2, P24). P10 (poet and essayist) described this as, “\textit{It’s hard for me to trace how money being made is based on the training data.}” Others felt it was hard to feel personally slighted when their contribution is so very small (P29, fan fiction writer). Still others were unsure exactly how compensation would even work: “\textit{It's not like a Spotify thing where you can be like, Okay, they listen to this track, or they read this chapter}” (P17). 

Despite many writers wanting compensation for the use of their work, they acknowledged that such compensation is likely to be small given the amount of data needed to train LLMs. P15, who has worked with LLMs in their creative practice, reflecting that they may be one of one million authors, noted that “\textit{I don't actually think that the companies who have created these models owe \textbf{me} a large amount of money}” even if the amount of money needed to compensate all authors would be significant. Other writers noted that they wouldn’t even necessarily want a small amount of compensation: “\textit{If it was gonna be 23 cents, because that's really how much it's worth, in some sense like… I think I should care more. But I don't actually want a check for 23 cents}” (P7). P32 noted that a very small amount of compensation in fact shouldn't be considered compensation at all---compensation necessitates that it compensates the writer, and a small amount wouldn’t actually be compensation for their work.

\supersubsection{Alternatives to compensation may also be impractical.} Some writers, either in addition to or in lieu of compensation, wanted to be credited for their contribution to an LLM (P14, P17, P18, P23, P24, P27). But
P24 noted that while being credited in a model was an interesting idea, it ultimately seemed futile: 

% if there is kind of a a prioritization system, or you know, if if it is, you know, more influenced by these ones than these ones, I think. Yeah, definitely highlighting ones that you kind of with sort of the the more influential ones to the work. P18
% Yeah, I think a public record would be really nice just to just so, I know. And other people know P17

\begin{quote}
    \textit{I guess, because no one will go look at [a list of contributing authors]. They won't. They won't care. If I'm going to write something, I just want to use the model. I just want to chat, write in my prompt and get my results. I'm not going to take the time to go, Gee! I wonder what 864,000 people were used to train this model.} (P24)
\end{quote}

A few writers did consider  ways to deal with the scale of LLM training data. P18 (fan fiction writer) wanted credit to be based on the output, such that outputs that were "\textit{more influenced}" by some writing could highlight the particularly relevant contributions. P15 suggested that rather than consider compensation at the individual level, “\textit{the moral thing to do would be to figure out some way to offset the disruption you're causing. The way that you would do that isn't necessarily just by paying everybody whose work has been fed into you know… it would be some other grant or taxation.}” Similarly P33 noted that industry disruption felt more salient than disruption to their individual career. P14 suggested that something like stock options would allow them to be compensated in line with the value that was created.
% In contrast, P31 argued that stock options were not compensation but more like being involved as a collaborator in a project.

\supersubsection{The impracticality of respect made some disengage entirely.} Writers’ understanding of the scale of LLM training data made it difficult for them to square their desire for respect as laborers and creatives. Writers wanted respect for their work, whether this be through consent, compensation, credit, or involvement, but felt that all and any of those were impractical. It would be impractical to get everyone’s consent (P11), compensation would be too small (P7, P15, P32), credit would be futile (P24), and no one expected to be invited into the upper ranks of technology corporations. This made many writers turn away from LLMs altogether; several of our interviewees actively disengaged with LLMs because of these issues, including never having tried them out themselves or avoiding learning more about them, even if they thought LLMs were theoretically or morally fine.

\section{Discussion}

\begin{quote}
    \textit{There is a deep sort of sadness to this notion that language can be mimicked in this way and that it can be sort of produced and really have no essential meaning or a central humanity, or at least not a detectable one. And then the idea that eventually these models and these applications will be able to simulate essential humanity. Or actually write moving, tear-jerking, compelling, bestselling novels. I mean, that's a pretty interesting idea. And I'm sure it's a place that we're going, and I'm not sure that we're prepared for it.} (P12)
\end{quote}

As seen in the above quote, writers often traversed a range of emotions in response to the implications of LLMs, and many writers noted that they were not sure what should be done. Overall, while writers had concerns about the implications of language models abstractly, most writers expressed distaste for large corporations using their work for dubious or unclear ends, and making a profit along the way.

% In our discussion, we first turn to the issue of power imbalances and social norms. Then we consider research that has tried to envision more responsible data collection practices. Finally, we outline how LLMs disrupt existing notions of the creative chain, discuss how LLMs could look more like libraries (which have a well understood and well respected place in the creative chain), and ways in which they may fail to do so.
% Finally, we propose future research to support these visions.

% \subsection{Power Differentials and Social Norms}

An implicit theme in many of our interviews was the way in which LLMs turned writing into data. Writers consider their work---writing---to be valued in a variety of ways (e.g. appreciation, compensation, prestige) by a human readership.
% : through the social norms of fan fiction, the purchase of original fiction, or the cultural practices of awards or licensing deals. 
LLMs, in defiance of these norms, subsume writing into the larger and more contentious category of data, where value becomes murky and the audience becomes algorithmic rather than human. Data has been analogized to labor \cite{arrieta2018should}, property \cite{prainsack2019logged}, and natural resource (e.g. "data mining"). Such analogies, although never perfect, attempt to sort through how data should be valued, especially data generated by, and often about, people. But is writing data? 

% Scholarly work on fan fiction is a rich area for understanding how usage norms can change as writing is transformed into, e.g., fan fiction. 

We argue that writing becomes data when it is used to train generative models, and this transformation creates the issues writers raise, many of which stem from a reorientation of their audience from people to algorithms or perhaps companies. Writers noted the power differential between themselves and the corporations making use of their work. This can be understood through the frame of precarious work \cite{allan2021PrecariousWork21st}; as writers feel that their work is increasingly precarious, the power differential between themselves and the organizations seeking to train LLMs grows larger.\footnote{Such power differentials have been explored in fan fiction communities. For instance, \citeauthor{busse2013remixing} (2013) report on fan fiction writers who note that the balance of power between two fan writers and between a fan writer and a professional writer is completely different, resulting in different social norms for remixing professional work than fan work.}

% Acknowledging this transformation lets us draw on a body of thinking about how to resist problematic uses of data. 
We can also see this dynamic as a form of extraction, in which writing is "taken" from writers for the profit of others.
% : writers struggled to imagine how they might be properly compensated given their personal insignificance in the face of the scale of language models and the power of their creators.
\citeauthor{couldry2019DataColonialismRethinking} (2019)  propose `data colonialism' as a new form of colonialism to make sense of the use of large amounts of data by a small group of corporate and government actors. 
If data colonialism is a useful framework to understand data collection for language models, which we think it may be, then we can turn to theories of decolonialism to understand how to resist it. \citeauthor{mohamed2020DecolonialAIDecolonial} (2020) put forth the idea of dismantling power assymmetries to resist data colonialism. This moves beyond our initial exploration of how data may or may not be collected in this context. 
% Decolonial tactics include a turn towards a critical technical practice (uncovering hidden assumptions and alternative ways of working), reverse tutelage (in which the periphery is considered expert), and political communities (e.g. grassroots organizations that can reform systems of hierarchy, knowledge, technology and culture) \cite{mohamed2020DecolonialAIDecolonial}. 

% Decolonial thinking points towards a larger understanding of the power differentials at stake when large amounts of data are collected by a few. 
While our findings often align with those studying visual artist communities---work by \citeauthor{jiangAIArtIts2023} (2023) finds that AI image generators create a chilling effect on cultural production, similar to our findings about industry impacts on writers---the direct economic impacts seem to be distinct (e.g. few of our interviewees expected to see personal economic impacts). However, the experience of precarious work and a lack of power remain the same. We believe that decolonial thinking can address a multitude of concerns (e.g. through pro-worker AI policies \cite{acemoglu2023CanWeHave} or public AI initiatives \cite{marda2024PublicAIMaking}) even as they differ in their specifics across communities. 
% Policy proposals have argued for 'pro-worker AI' \cite{acemoglu2023CanWeHave} which would harness AI capabilities to decrease precarious work, rather than increase it. Other organizations have argued for 'Public AI', a robust ecosystem that promotes "promote public goods, public orientation, and public use" \cite{marda2024PublicAIMaking}.
% We see value in writers forming political communities that can resist the power of large technology corporations, and provide expertise into how they do or do not want to contribute to the development of language models.

When asked who they expect to best advocate for their interests, writers did not think governments, publishers, universities, and certainly not technology companies, had their interests at heart. Some writers relied on their agents, but most writers would only trust other writers to safely and respectfully steer LLM creation. We recommend that writers form political communities which can take the helm in determining how their work is used.

% \subsection{Navigating the Role of Writing-as-Data in the Creative Chain}

As we reported in \subcode{the creative chain} (\ref{sec:creativechaing}), while writers expect their writing to be part of a larger cultural project over which they have little control,
% , and take pleasure in inspiring other writers or contributing to education and literacy projects. However, 
LLMs challenged their notion of what it meant to participate in the creative chain. 
% While typical engagement with their writing comes directly from other people, LLMs create an intermediary by turning writing into data. It is unclear how the user of an LLM engages with the data it was trained on.
Some writers envisioned how LLMs could be more aligned with the kinds of cultural projects that they wanted to contribute to. For instance, several writers discussed the positive values of libraries, and considered a world where LLMs were more like libraries. Yet, libraries and language models have important differences. The understood intent of libraries is that a wider variety of people can have direct access to your writing, whereas the final audience of an LLM may be ignorant of individual author contributions. Libraries are non-profit institutions that serve the general public. They are also physical locations that serve their local community in a variety of ways. 

Some of these differences could be addressed: LLMs could be created by non-profit institutions to serve the general public, as recommended in \cite{marda2024PublicAIMaking} and seen in projects like \cite{le2023bloom} and \cite{soldaini2024dolma}. The contributions of individual authors could be made more clear, for instance through work on attributing generated text to specific training data \cite{dengcomputational, hammoudeh2024training}. However, it is likely difficult to completely remove the obscurity that lies between training data and generated text, and this disjuncture may continue to cause unique problems.

Although our work primarily investigated training data for \textit{large} language models, writers did express interests in smaller models that were either private or local to a small community. As P2 put it, “\textit{I really love the idea of personal, more customized models for your friends and groups … it just feels like fun exploration.}” Such models, with smaller amounts of creative writing in the training data, would make their place in the creative chain more clear, and make it easier to engage in respect for those who contributed to the training data.

\section{Future Work}

We outline two avenues of future work. The first is how to prevent the use of writing as training data without consent. The second is how to create models that writers would want to contribute to. We see these as the negative and positive sides of futuring: how to discourage a future we don't want, and create a future we desire.

\textbf{Preventing nonconsensual usage.} While there is preliminary work on detecting if certain text has been used to train a model \cite{hammoudeh2024training}, or applying watermarks to text \cite{lauprotecting, qiang2023natural, yoo-etal-2023-robust}, more robust and easy-to-access methods would allow writers to determine whether their writing has been used without their consent, which is the first step in recourse. Legal protections could enforce this, but social norms could also discourage institutions from nonconsensual usage.\footnote{For example, fan fiction communities have been found to have highly consistent social norms which are policed internally \cite{fiesler2019CreativityCopyrightCloseKnit}.}
Research into the economic impacts of LLMs on writers could also support this agenda.

\textbf{Creating writer-controlled language models.} Community datasets could allow writers to contribute only to projects they believed in. Such work would treat writers as experts rather than resources, and could draw on research on community archives \cite{FlinnAndrew2009Wmwa,flinn2007community,huvila2008participatory,caswell2016suddenly} and archival practices for machine learning \cite{jo2020lessons,desai2024archival}. Technical work that supports smaller models that can be run locally, or models that can more easily interact with sensitive training data \cite{minSILOLanguageModels2023}, could support community models which may not have the compute power of corporate ones.

\section{Conclusion}

Most of the writers we talked to were not against the use of creative writing to train LLMs in theory, but rather against the way in which LLMs have been steered and created thus far. Even those who could not imagine why someone would want to use an LLM in their writing practice were open to the idea that people should be allowed to experiment with this technology. Still, most writers were either exasperated with the situation, seeing no way to influence it, or they simply accepted it, often due to their interpretation of their small contribution given the scale of LLM training data. In this work, we attempt to represent how creative writers are reasoning about the use of their writing as training data for LLMs, and articulate a path forward that gives writers the respect they deserve.

%TC:ignore

% \begin{acks}
% \end{acks}

\bibliographystyle{ACM-Reference-Format}
\bibliography{citations}

\appendix

\section{Self-Reported Demographic Data from Interviewed Writers}
\label{app:demographics}

We spoke with 33 writers who write across a variety of genres. The majority of writers we spoke with (22 out of 33) write in multiple genres, as indicated in \autoref{tab:genres}.

\begin{table}[h] 
    \centering 
    \caption{Writing Genre of Interviewees}
    \label{tab:genres}

    \begin{tabular}{l r} 
        \toprule
        \textbf{Genre} & \textbf{} \\ 
        \midrule 
        \textit{Fiction} & \textit{22} \\
        \quad \small Literary Fiction & \small 10 \\
        \quad \small Genre Fiction & \small 9 \\
        \quad \quad \scriptsize{Sci-Fi and Fantasy} & \scriptsize{5} \\
        \quad \quad \scriptsize{Romance} & \scriptsize{3} \\
        \quad \small Fan Fiction & \small 5 \\
        \quad \small Young Adult (YA) Fiction & \small 3 \\
        & \\ \
        \textit{Non-Fiction} & \textit{14} \\
        \quad \small Personal Essay and Memoir & \small 11 \\
        \quad \small Journalism & \small 5 \\
        & \\ \
        \textit{Poetry} & \textit{15} \\
        \bottomrule 
    \end{tabular}
\end{table}

The writers we spoke with also share their work via a variety of publishing or delivery methods, detailed in \autoref{tab:distrib}. As with genre, the majority of writers we spoke with (18 out of 33) also share their writing through multiple venues and means.

\begin{table}[h] 
    \centering 
    \caption{Method of Delivery for Interviewees}
    \label{tab:distrib}

    \begin{tabular}{l r} 
        \toprule
        \textbf{Publishing or Delivery Method} & \textbf{} \\ 
        \midrule 
        \textit{Traditional} & \textit{26} \\
        \quad \small Major Publisher ("Big 5") & \small 9 \\
        \quad \small Independent Publishers, Literary Magazines, and Journals & \small 21 \\
        & \\ \
        \textit{Self-Published} & \textit{15} \\
        \quad \small Blogs and Forums & \small 9 \\
        \quad \small Kindle Direct or equivalent & \small 6 \\
        & \\ \
        \textit{Performance} & \textit{7} \\
        \quad \small Readings & \small 5 \\
        \quad \small Installations and Exhibits & \small 3 \\
        \bottomrule 
    \end{tabular}
\end{table}

The degree to which writers were compensated for their creative work varied significantly, detailed in \autoref{tab:compensation}.

\begin{table}[h] 
    \centering 
    \caption{Degree of Professionalization for Interviewees}
    \label{tab:compensation}

    \begin{tabular}{l r} 
        \toprule
        \textbf{Compensation for Creative Work} & \textbf{} \\ 
        \midrule 
        {Most income comes from creative writing practice} & {11} \\
        {Most income comes from writing-related profession} & {10} \\
        {Most income does not come from writing} & {8} \\
        {Unknown} & {4} \\
        \bottomrule 
    \end{tabular}
\end{table}

We also noted how writers self-reported their use of LLMs; see \autoref{tab:engagement}. The vast majority (23 out of 33) were "dabblers"—those who had tried out LLMs, but do not use them in a regular or intensive manner. A significant minority (8 out of 33) use LLMs regularly in either their paratextual or central creative writing practices. 

\renewcommand{\arraystretch}{1.25}
\begin{table}[h] 
    \centering 
    \caption{Engagement with LLMs in Writing Practice}
    \label{tab:engagement}
    \begin{tabular}{p{7cm} r} 
        \toprule
        \textbf{Engagement with LLMs in Writing Practice} & \textbf{} \\ 
        \midrule 
        {None} & {3} \\
        {Some (e.g., "trying it out" or "dabbling")} & {23} \\
        {Regular use in peripheral writing and publishing activities (e.g., marketing, paratextual use)} & {3} \\
        {Regular use in creative composition (e.g., brainstorming, narrative outline, turns of phrase, etc.)} & {5} \\
        \bottomrule 
    \end{tabular}
\end{table}

\section{Interview Guideline}
\label{app:interviewguideline}

The following guideline represents our final interview guideline. It includes core questions (solid bullet points) and potential follow-up questions (dashed bullet points). The guideline acted as a structure for the interview; not all interviews went through all follow-up questions or went through the questions in this precise order, depending on how the conversation was progressing. Note that two questions use a stimulus, which was either verbally or visually presented to interviewees to aid in the conversation.

\subsection{General writing questions}

\begin{itemize}
    \item Could you briefly describe your writing education or journey?
    
    \begin{itemize}
        \item What initially inspired you to become a writer? Can you recall a specific moment or influence that ignited your passion for writing?

        \item How many years have you been writing as a job? 

        \item How has your writing style or focus evolved over the course of your career? 
    \end{itemize}

    \item Could you briefly describe your current writing practice?

    \begin{itemize}
        \item What is your typical writing process? 

        \item What genre or type of writer do you identify with most? 

        \item How do you seek and incorporate feedback on your work? Are there specific individuals or groups you rely on for constructive criticism? 

        \item Is writing your main profession, or do you have other jobs? How do you divide your time? 
    \end{itemize}

    \item Do you share or publish your writing?

    \begin{itemize}
        \item What is your approach to sharing or publishing your work? Who is involved in the process? 

        \item Has it changed over time? How?
    \end{itemize}
\end{itemize}

\subsection{Knowledge of and attitudes towards LLMs}

\begin{itemize}
    \item How much do you know about large language models like ChatGPT or Gemini (formerly Bard)?

    \begin{itemize}
        \item How would you describe LLMs to someone who doesn’t know what it is? 
    \end{itemize}

    \item Have you ever used these models?

    \begin{itemize}
        \item What did you do or try?

        \item What did you like or dislike about them?

        \item Are there particular applications (perhaps that you see others engaged in) you like or dislike?
    \end{itemize}

    \item Have you ever used these in your writing process? In what ways? 

    \begin{itemize}
        \item If yes: Has your opinion on LLMs now changed since you first used one? 

        \item What do you like or not like about using them? Why?

        \item Either yes/no: Are there specific aspects of your writing practice where you see these models being relevant or useful?
    \end{itemize}

    \item Are you interested in using them in the future?

    \begin{itemize}
        \item If yes: in what stage of the writing process would you use them? 

        \item Are there applications that you particularly like or dislike? 

        \item If no: why not?
    \end{itemize}

    \item Have you thought much about the training data for these models? 

    \begin{itemize}
        \item What are your thoughts on the source of the training data for these models?
    \end{itemize}
\end{itemize}

\subsection{Knowledge of and attitudes towards language model training data}

\begin{itemize}
    \item Do you know if any of your writing was used to train a large language model? 

    \begin{itemize}
        \item If yes: How did you find out?

        \item If don't know: Have you wondered? Why or why not?
    \end{itemize}

    \item How do you feel about the actual, potential, or hypothetical inclusion of your writing in an existing language model?

    \begin{itemize}
        \item Are there details about the use of the model that matter to you?
        \item \textbf{Note: Use stimulus 1}
    \end{itemize}

    \item Under what conditions, if any, would you want or consent to your own writing be included in an LLM training dataset?

    \begin{itemize}
        \item Why do you prefer [x] over [y]? 
        \item \textbf{Note: Use stimulus 2}
        \item Can you walk me through your thought process in your reactions? 

        \item If all your conditions were met, would you want your writing included? Why or why not?

        \item If compensation is desired: What if compensation is very low? What about other forms of compensation, like stock options, or money going towards professional organizations or scholarships?

    \end{itemize}

    \item How might these conditions differ for different kinds of writing? e.g. Published v. unpublished work, fiction v. nonfiction. Why?

    \item What material ways can you imagine your writing career being affected by LLMs?

    \begin{itemize}
        \item Is there anything about the use of your writing data in particular that contributes to these material effects? (Versus LLMs generally.)
    \end{itemize}

    \item Are there any upsides of your work being included? e.g. For people to search, summarize, get recommendations.

    \item What institutions do you think should be collecting and protecting writing data? Or collecting opt-in/opt-out information? e.g. Government labs, universities, professional organizations like the Author’s Guild, non-profits.

    \item Who do you believe would advocate best for your interests? e.g. agents, publishers, lawyers, non-profits.

    \begin{itemize}
        \item Who do you trust to advocate for your interests?

        \item Who do you not trust?
    \end{itemize}

    \item If you were to sign a contract with LLM provisions, how would you negotiate?

\end{itemize}

\subsection{Stimulus 1: Hypothetical usage scenarios}

Commercial interests of the organization that developed the model:

\begin{itemize}
    \item You are notified that several of your books were used by an LLM developed by a for-profit company. Your books are unavailable online for free. The company is charging a monthly subscription for access to their model.
    \item Contrast A (research purposes): The LLM was developed by a university or government lab for research purposes only (i.e. non-commercial).
    \item Contrast B (non-profit): The LLM was developed by a non-profit dedicated to providing LLM access globally for free.
    \item Contrast C (writing was available online): Instead of your books, the LLM used a series of short stories you wrote which are available online for free, although you were paid for these stories by the publisher and signed a contract saying you still own the rights to the stories.
\end{itemize}

How people are using the model:

\begin{itemize}
    \item You are notified that several of your books were used to train an LLM. People are using this LLM primarily to help with business-focused writing tasks like writing cover letters, business emails, or press releases.
    \item Contrast A (creative writing): People are using the LLM for creative writing like writing stories, personal essays, or poems.
    \item Contrast B (pro-social applications): People are using the LLM in professional contexts like lawyers using it to streamline support in pro bono cases, doctors using it to support diagnosis of rare diseases, teachers using it to help teach difficult concepts (like chemistry) or scientists to make new discoveries.
\end{itemize}

\subsection{Stimulus 2: Conditions of consent}

Compensation:

\begin{itemize}
    \item fixed, one-off payment for a specific contribution
    \item ongoing compensation or royalties entail recurring payments based on the use of the model (i.e. like royalties for book sales)
    \item ongoing compensation based on usage of the model related to your work; e.g. if someone uses a model for their math homework, you don’t get compensated, but if it’s used for creative writing, you do get compensated.
\end{itemize}

Credit:

\begin{itemize}
    \item public record of contributors listed on the LLM platform
    \item Incorporate a mechanism for the model to acknowledge the source of its training data when generating outputs. (i.e. citation) 
    \item List contributors as collaborative authors on associated publications or reports

\end{itemize}

Use of model:

\begin{itemize}
    \item usage only in education settings
    \item disallow usage that would compete with your interests (e.g. disallow use to write novels that could theoretically compete with your novel)
\end{itemize}

Commercial:

\begin{itemize}
    \item usage only in non-commercial settings (e.g. research, art, or non-profit)
    \item require compensation in commercial applications

\end{itemize}

Personal access to model:

\begin{itemize}
    \item Exclusive access to the model 
    \item free access to the model

\end{itemize}

%TC:endignore

\end{document}